\documentclass{aa}  
\usepackage{graphicx}
\usepackage{bm}
\usepackage{txfonts}
\usepackage{float}
\usepackage{xcolor}
\usepackage{natbib,twoopt}
\usepackage[breaklinks=true]{hyperref} %% to avoid \citeads line fills
\bibpunct{(}{)}{;}{a}{}{,} %% natbib format for A&A and ApJ
\makeatletter
\newcommandtwoopt{\citeads}[3][][]{\href{http://adsabs.harvard.edu/abs/#3}%
	{\def\hyper@linkstart##1##2{}%
		\let\hyper@linkend\@empty\citealp[#1][#2]{#3}}}
\newcommandtwoopt{\citepads}[3][][]{\href{http://adsabs.harvard.edu/abs/#3}%
	{\def\hyper@linkstart##1##2{}%
		\let\hyper@linkend\@empty\citep[#1][#2]{#3}}}
\newcommandtwoopt{\citetads}[3][][]{\href{http://adsabs.harvard.edu/abs/#3}%
	{\def\hyper@linkstart##1##2{}%
		\let\hyper@linkend\@empty\citet[#1][#2]{#3}}}
\newcommandtwoopt{\citeyearads}[3][][]%
{\href{http://adsabs.harvard.edu/abs/#3}
	{\def\hyper@linkstart##1##2{}%
		\let\hyper@linkend\@empty\citeyear[#1][#2]{#3}}}
\makeatother

\usepackage[normalem]{ulem}
\usepackage{soul}

\begin{document}

	\title{Magnetic seed generation by plasma heat flux in accretion disks} 
	
	\author{N. Villarroel-Sep\'{u}lveda
		\inst{1}
		\and
		F. A. Asenjo \inst{2}
		\and 
		P. S. Moya \inst{1} 
	}
	\institute{Departmento de F\'{\i}sica, Facultad de Ciencias,
		Universidad de Chile, Las Palmeras 3425, 7800003, Ñuñoa, Santiago, Chile\\
		\email{nicolas.villarroel@ug.uchile.cl, pablo.moya@uchile.cl}
		\and
		Facultad de Ingeniería y Ciencias, Universidad Adolfo Ibáñez, 7941169 Peñalolén, Santiago, Chile
		\email{felipe.asenjo@uai.cl} 
	}
	
	\date{\today}% It is always \today, today,
	
	\abstract
	{Magnetic batteries are potential sources that may drive the generation of a seed magnetic field, even if this field is initially zero. These batteries can be the result of non-aligned thermodynamic gradients in a plasma, as well as of special and general relativistic effects. So far, magnetic batteries have only been studied in ideal magnetized fluids.}
	{We study the non-ideal fluid effects introduced by the energy flux in  the vortical dynamics of a magnetized plasma in curved spacetime. We propose a novel mechanism for generating a heat flux-driven magnetic seed within a simple accretion disk model around a Schwarzschild black hole.} 
	{We use the 3+1 formalism for the splitting of the space-time metric into space-like and time-like components. We study the vortical dynamics of a magnetized fluid with a heat flux in the Schwarzschild geometry in which thermodynamic and hydrodynamic quantities are only dependent on the radial coordinate. Assuming that the magnetic field is initially zero, we estimate linear time evolution of the magnetic field due to the inclusion of non-ideal fluid effects. } 
	{When the thermodynamic and hydrodynamic quantities vary only radially, the effect of the coupling between the heat flux, spacetime curvature and  fluid velocity acts as the primary driver for an initial linearly time growing  magnetic field. The plasma heat flux completely dominates the magnetic field generation at an specific distance from the black hole, where the fluid vorticity vanishes. This distance depends on the thermodynamical properties of the Keplerian plasma accretion disk. These properties control the strength of the non-ideal effects in the generation of seed magnetic fields.}
	{We find that heat flux is the main driver of a seed magnetic field in black hole accretion disks if the geometry, plasma dynamics and thermodynamics share the same axial symmetry. This suggests that non-ideal fluid effects may play a major role in the magnetization of astrophysical plasmas.}

	\keywords{ accretion, accretion disks --
		magnetic fields --
		plasmas
	}
	
	\maketitle

	\section{Introduction}
	
	Large-scale magnetic fields appear to be ubiquitous in the known physical universe. Dynamo mechanisms can significantly amplify a small magnetic field but cannot do so if the field is initially absent. This has led to growing interest in exploring new methods for creating small magnetic fields that can act as seeds for these large-scale structures. In recent years, magnetic seed generation through mechanisms such as the Weibel instability \citep{Zhang2020} and the Biermann battery \citep{Pilgram_2022} have been successfully observed in laboratory plasmas.
	
	In this article, we explore the generation of a seed magnetic field through a heat flux-induced battery mechanism in the plasma surrounding a spherically symmetric, static black hole. We will consider a non-gravitating plasma fluid with heat flux, which evolves in a background curved spacetime defined by the Schwarzschild metric. Our treatment of this system follows the framework of the unified magnetofluid theory \citep{mahajan_2003}, which describes the coupling between plasma hydrodynamics, thermodynamics, and electrodynamics in curved spacetime. We also adopt the 3+1 formalism for the foliation of spacetime \citep{Arnowitt_Deser_Misner_2008}, that allows the separation of four-dimensional vectorial and tensorial quantities into space-like and time-like components in the reference frame of an Eulerian (or fiducial) observer \citep{gourgoulhon_2012}.  
	
	In this case, its stress-energy tensor $T^{\mu\nu}$ is \citep{misner_thorne_wheeler_kaiser_2017} 
	\begin{subequations}
		\begin{align}
			T^{\mu\nu}=&\frac{h}{c^2}U^{\mu}U^{\nu}+pg^{\mu\nu}+T_h^{\mu\nu},\label{eq:stress-energy_a}\\
			T_h^{\mu\nu}=&\frac{1}{c^2}\left(q^{\mu}U^{\nu}+q^{\nu}U^{\mu}\right),\label{eq:stress-energy_b}
		\end{align}
	\end{subequations}
	
	where $T_h^{\mu\nu}$ accounts for the energy transfer within the fluid due to the heat flux vector $q^{\mu}$. Here, $c$ is the speed of light,
	$h=\rho+p$ is related to the fluid's enthalpy density, $\rho$ is the mass-energy density, $p$ is the thermodynamic pressure, and $g^{\mu\nu}$ is the metric of the particular geometry. The plasma fluid four-velocity is $U^{\mu}=\Gamma(c,\vec{v})$, with $\vec v$ the plasma three-velocity, and 
	$\Gamma$ is the general-relativistic Lorentz factor. For the sake of simplicity, the heat flux term is the only considered non-ideal component of the stress-energy plasma tensor.
	Besides, the plasma fluid obeys the continuity equation $\nabla_{\mu}(nU^{\mu})=0$, where $n$ is the plasma density, and $\nabla_\mu$ is a covariant derivative. The whole system is coupled to the curved spacetime Maxwell equations $\nabla_{\nu}F^{\mu\nu}=({4\pi}/{c})J^{\mu}$ and $\nabla_{\nu}(\star F^{\mu\nu})=0$, where $F^{\mu\nu}$ is the electromagnetic tensor and $\star F^{\mu\nu}=({1}/{2})\epsilon^{\mu\nu\rho\sigma}F_{\rho\sigma}$ is its Hodge dual, and $J^{\mu}=qnU^{\mu}$ is the fluid's four current, with $q$ being the charge of the plasma species. 
	
	The equation of motion governing the plasma dynamics is
	\begin{align}
		\nabla_{\nu}T^{\mu\nu}&=\frac{qn}{c}U_{\nu}F^{\mu\nu}. \label{eq:Lorentz}
	\end{align}
	By introducing the quantity $f=h/mn$, we define the fully antisymmetric tensor $S^{\mu\nu}=[\nabla^{\mu}(fU^{\nu})-\nabla^{\nu}(fU^{\mu})]/c$. Then, using it in Eq.~(\ref{eq:Lorentz}), we obtain \cite{mahajan_2003,asenjo_mahajan_qadir_2013}
	\begin{align}
		\frac{1}{c}U_{\nu}M^{\mu\nu}&=-\frac{T}{q}\nabla^{\mu}\sigma + \frac{\nabla_{\nu}T_h^{\mu\nu}}{qn}, \label{eq:magnetofluid}
	\end{align}
	where $M^{\mu\nu}=F^{\mu\nu}+(m/q)S^{\mu\nu}$ is the fully antisymmetric magnetofluid tensor \cite{mahajan_2003}, and $\sigma$ is the fluid entropy per baryon obtained from the first law of thermodynamics $\mathbf{d}\rho=(h/n)\mathbf{d}n+nT\mathbf{d}\sigma$. The inclusion of the analogous electromagnetic tensor $S^{\mu\nu}$, resulting from the anti-symmetric gradient of the flux $fU^{\mu}$, is not arbitrary. This quantity accounts for the coupling between the magnetic vector potential $A^{\mu}$ and the fluid's momentum through the generalized momentum $\Pi^{\mu}=mU^{\mu}+\left({q}/{c}\right) A^{\mu}$, which results in a common equation shared by the time evolution of the magnetic field and fluid vorticity in the classical derivation of the Biermann battery \citep{Biermann_Schlüter_1950, Kulsrud_2005}.

		It is worth mentioning that the mathematical framework of the unified magnetofluid formalism is not unique when it comes to the study of coupling between hydrodynamics, electrodynamics and general relativity. \cite{Khanna_1998a} derived the magnetohydrodynamic equations of a two-component plasma in an axisymmetric motion around a Kerr black hole employing the 3+1 formalism. Following this work, \cite{Khanna_1998b} showed that gravitational contributions to the partial pressure of electrons are able to sustain currents that work as a gravitomagnetic Biermann battery in Kerr spacetime.
		A different approach was taken by \cite{Contopoulos_1998}, who proposed Poynting-Robertson radiation drag as a source for an azimuthal current that induces poloidal magnetic fields in an electron-ion plasma.  This mechanism was dubbed \textit{Cosmic Battery} and has been discussed extensively in the context of Kerr black hole accretion disks, including general-relativistic effects \citep{Koutsantoniou_2014, Sadowski_2017, Contopoulos_2018}.
	
	\section{Equation for general-relativistic vorticity}
	
	Eq.~\eqref{eq:magnetofluid} is the starting point to obtain the seed magnetic field from the contribution of the plasma heat flux.
	For this, we analyze this equation in the background Schwarzschild spacetime. We use a 3+1 decomposition of this metric in terms of ADM variables. These are the lapse function $\alpha$, the future-oriented unit four-vector as ${n}^\mu=(1/\alpha,0,0,0)$, and the space-like projector, $\gamma_{\mu\nu}=g_{\mu\nu}+n_{\mu\nu},$ such that $n^\mu \gamma_{\mu\nu}=0$. 
	The components of the metric can be inferred from the Schwarzschild line element
	$ds^2=-\alpha(r)^2dt^2+\alpha(r)^{-2}dr^2 + r^2d\Omega^2$, where $\Omega$ is the spherical solid angle. Since the Schwarzschild spacetime possesses no extrinsic curvature, the gradient of the unit time-oriented dual 1-form is given by $\nabla_{\nu}n_{\mu}=-n_{\nu}a_{\mu}$,
	where $a_{\mu}=n^{\nu}\nabla_{\nu}n_{\mu}=\partial_{\mu}\alpha/\alpha$.
	The above formalism allows the decomposition of physical quantities 
	as $U^{\mu}=c\Gamma(\alpha n^{\mu}+\gamma^{\mu}{}_{i}v^i/c)$, and $M^{\mu\nu}=n^{\mu}\xi^{\nu}-n^{\nu}\xi^{\mu}-\epsilon^{\mu\nu\rho\sigma}\Omega_{\rho}n_{\sigma}$ \cite{asenjo_mahajan_qadir_2013}.  The components
	\begin{align}
		\xi^\mu&=n_\nu M^{\mu\nu},\\
		\Omega^\mu&=\frac{1}{2}\epsilon^{\mu\alpha\beta\gamma}n_\alpha M_{\beta\gamma},
	\end{align}
	represent generalized electric-like and magnetic-like fields, respectively, in the unified magnetofluid formalism, the latter being referred to as the plasma's generalized vorticity. In vectorial notation, these space-like four-vectors are given by
	\begin{align}
		\bm{\xi}&=\mathbf{E}-(m/\alpha q)[\bm{\nabla}(f\alpha^2\Gamma)+\partial_t (f\Gamma\mathbf{v})],\\
		\mathbf{\Omega}&=\mathbf{B}+({m}/{q})\bm{\nabla}\times\left(f\Gamma \mathbf{v}/c\right),
	\end{align}
	where ${\bf E}$ and $\mathbf{B}$ are the electric and magnetic fields, respectively \cite{asenjo_mahajan_qadir_2013}.
	\color{black}
	
	Using the above $3+1$ decomposition, we can calculate the space-like projection of Eq.~(\ref{eq:magnetofluid})  by contraction with $\gamma^{i}{}_{\mu}$. Thus, we obtain
	\begin{align}
		\alpha\xi^i+n_{\sigma}\epsilon^{\sigma ijk}v_j\Omega_k=-\frac{T}{q\Gamma}\nabla^{i}\sigma + \frac{1}{qn\Gamma}\gamma^{i}{}_{\mu}\nabla_{\nu}T_h^{\mu\nu}. \label{eq:gen-ef}
	\end{align}
	Here, the gradients $\nabla^{i}$ are used interchangeably with the space-like projection of the four-gradient $\gamma^i{}_{\mu}\nabla^{\mu}$, since for a non-rotating geometry $n^i=0$, and therefore $\gamma^{i}{}_{\mu}=g^{i}{}_{\mu}$.
	
	The last term on the right-hand side contains new, non-ideal effects on describing the generalized vorticity dynamics. Explicit calculation of it, using the definitions in \eqref{eq:stress-energy_a} and \ref{eq:stress-energy_b}, yields  $-({1}/{qn\Gamma})\gamma^{i}{}_{\mu}\nabla_{\nu}T_h^{\mu\nu}=\Lambda^i$, with
	\begin{align}
		\Lambda^i
		=&\frac{1}{q\Gamma c^2}\gamma^i{}_{\mu}U^{\nu}\nabla_{\nu}(q^{\mu}/n)+
		\frac{v^i}{qnc^2}\nabla_{\mu}q^{\mu}+\frac{1}{qn\Gamma c^2}\gamma^{i}{}_{\mu}q^{\nu}\nabla_{\nu}U^{\mu}.
		\label{eq:lambdaheatflux}
	\end{align}
	
	To find the dynamical behavior of the general-relativistic vorticity, we need to use the properties of $M^{\mu\nu}$. The vorticity equation can be obtained from the constraint
	$\nabla_{\nu}\star {M}^{\mu\nu}=0$, 
	where $\star{M}^{\mu\nu}=\Omega^{\mu}n^{\nu}-\Omega^{\nu}n^{\mu}-\epsilon^{\mu\nu\rho\sigma}\xi_{\rho}n_{\sigma}$ is the Hodge dual of the magnetofluid tensor.
	Projecting this equation onto $n_{\mu}$ results in the generalized magnetic Gauss's law $(\nabla_{\mu}-a_{\mu})\Omega^{\mu}=0$. On the other hand,
	its space-like projection  becomes
	\begin{align} 
		\frac{1}{\alpha(r)}\frac{\partial \Omega^i}{\partial t}=-\frac{c\alpha(r)}{\sqrt{|g|}}\varepsilon^{0ijk}(\nabla_j+a_j)\xi_{k},\label{eq:time_omega} 
	\end{align} 
	which is completely equivalent to the analogous equation found in Ref.~\cite{asenjo_mahajan_qadir_2013}. Here, we have related the zero-gradient ($\nabla \epsilon =0$) covariant Riemann volume form $\epsilon$ to the Levi-Civita symbol $\varepsilon$, as $\epsilon^{\mu\nu\rho\sigma}={\textnormal{sgn}(g)}\varepsilon^{\mu\nu\rho\sigma}/{\sqrt{|g|}}$. 
	Thereby, we obtain the vorticity equation by replacing Eq.~(\ref{eq:gen-ef}) in Eq.~(\ref{eq:time_omega}), to obtain 
	\begin{align}
		&\frac{\partial \Omega^i}{\partial t}-c\varepsilon^{0ijk}\varepsilon_{0k\ell m}\nabla_jv^{\ell}\Omega^{m}=\frac{c\alpha(r)}{\sqrt{|g|}}\varepsilon^{0ijk}\nabla_j\Big[\frac{T\nabla_{k}\sigma}{q\Gamma}+\Lambda_{k}\Big].\label{eq:vorticity}
	\end{align}

	In vector notation, Eq.~(\ref{eq:vorticity}) is equivalent to ${\partial \vec{\Omega}}/{\partial t}-\vec{\nabla}\times(\vec{v}\times\vec{\Omega})=\vec{\nabla}\times({T\vec{\nabla}\sigma}/({q\Gamma})+\vec{\Lambda}\big)$, where the three-vectors and three-gradients are obtained by taking the space-like projection of the corresponding four-vectors and covariant derivatives, respectively.  The previous expression contains three feasible magnetic seed drives. The first one is the relativistic baroclinic drive (Biermann battery)  corrected by spacetime curvature $\vec \Xi_B=\vec{\nabla} T\times\vec{\nabla}\sigma/(q\Gamma)$, which can produce a seed magnetic field under particular configurations of the plasma's thermodynamic properties in Schwarzschild spacetime \citep{asenjo_mahajan_qadir_2013}. The second one is the relativistic drive $\vec \Xi_R=-T\, \vec{\nabla} \Gamma\times\vec{\nabla}\sigma/(q\Gamma^2)$ \citep{Mahajan_Yoshida_2010,asenjo_mahajan_qadir_2013}, which can generate magnetic fields through the interaction of relativity and plasma thermodynamics. The last one, the heat flux drive
	\begin{equation}
		\vec{\Xi}_q=\vec{\nabla}\times\vec{\Lambda},
		\label{heatfluxdrive}
	\end{equation}
	is the main result of this work. This nonlinear drive, being the curl of vector \eqref{eq:lambdaheatflux}, has a completely different nature compared to the other ones. Below, we show that in the most simple scenario for a plasma in an accretion disk around a Schwarzschild black hole, the only possibility of generating a seed magnetic field is the action of the heat flux drive.

	\section{Vorticity seed generation by heat flux in accretion disks}
	
	Consider a plasma in a thin accretion disk (in equatorial plane $\theta=\pi/2$) around a Schwarzschild black hole. In this case, the lapse function only depends on the radial coordinate $\alpha(r)=\big(1-{r_S}/{r}\big)^{1/2}$, where $r_S={2GM}/{c^2}$ is the Schwarzschild radius,  with $M$ the black hole mass and $G$  the universal gravitational constant. Also, $\gamma_{ij}={\alpha^{-2}}\delta^{r}{}_{i}\delta^{r}{}_{j}+r^2[\delta^{\theta}{}_{i}\delta^{\theta}{}_{j}+\sin^2\theta\delta^{\phi}{}_{i}\delta^{\phi}{}_{j}]$.
	
	The simplest scenario consists of every single plasma variable depending only on the radial Schwarzschild coordinate. In this case, it is straightforward to realize that the baroclinic drive $\vec\Xi_B=0$, and the relativistic drive $\vec\Xi_R=0$ (as all gradients are parallel). Only the heat flux drive \eqref{heatfluxdrive} does not vanish under this simple assumption (as shown below). Of course, one can invoke more sophisticated dynamical plasma scenarios where $\vec\Xi_B$ or $\vec \Xi_R$ do not vanish in accretion disks.
	For example, in Ref.~\cite{asenjo_mahajan_qadir_2013} a non-radial plasma temperature profile was assumed, or in Ref.~\cite{bhattacharjee_das_mahajan_20152}
	the orbit of the particles in the accretion plasma disk was not stable, but inward spirals into the event horizon. In those cases, the heat flux drive will also contribute.

	Although the plasma orbits in accretion disks are not stable in general \citep{Vietri_2008}, this process is expected to occur very slowly. Therefore, for this work, we consider the plasma to be in stable Keplerian orbits in the black hole equatorial plane. Also, we resume to the simplest assumption that all plasma quantities depend on radial direction only, thus that the baroclinic and relativistic drives vanish identically. Thereby, we consider only Keplerian orbits with an angular velocity $v^{\phi}=\dot{\phi}=c\sqrt{r_S/2r^3}$. 
	Since $v^2\approx (rv^{\phi})^2$, we obtain the Lorentz factor $ \Gamma(r)=(1-{r_{ph}}/{r})^{-1/2}$,
	which allows any orbit beyond $r\geq r_{ph}$, where $r_{ph}=3r_S/2$ is the radius of the photosphere, describing the innermost unstable circular orbit for massless particles.
	
	In order to evaluate the heat flux drive, we first assume that the heat flux one-form $q^\mu$ has components given by the general-relativistic Tolman's law \citep{misner_thorne_wheeler_kaiser_2017}
	\begin{subequations}
		\begin{align}
			q_0&=0,\label{eq:heat_a}\\
			q_j&=-\frac{\kappa}{\alpha}\partial_j(\alpha T),\label{eq:heat_b}
		\end{align}
	\end{subequations}
	
	where the temperature-dependent collision-driven thermal conductivity coefficient is given by the \cite{Spitzer_Härm_1953} relation $\kappa(T)=\kappa_0(T/T_0)^{5/2}$, where $\kappa_0$ and $T_0$ are dimensional constant yet to be determined. We note that the thermal conduction coefficient $\kappa_0$ has been normalized to the referential temperature $T_0$, and therefore has cgs units of $[\textnormal{g}\cdot\textnormal{cm}\cdot\textnormal{K}^{-1}\cdot\textnormal{s}^{-3}]$, which are those of the thermal conduction coefficient in Fourier's law of heat conduction. A conductive heat flux of this kind has often been utilized in the literature regarding general relativistic plasmas and fluids \citep{misner_thorne_wheeler_kaiser_2017, Tanaka_2006, Chandra_2015, Schobert_2019}. 
	
	The temperature is considered to be a product of the accretion disk's blackbody radiation;  $T=(k(r)/\sigma_{SB})^{1/4}$, where $\sigma_{SB}$ is the Stefan-Boltzmann constant and $k(r)$ is the time-averaged radiation flux \citep{Thorne_1974}, which is given by $k(r)=-({\Dot{M}}/{4\pi r})({\partial_{r}v^{\phi}}/{(\tilde{E}-v^{\phi}\tilde{L})^2})\int_{r_{I}}^{r}(\tilde{E}-v^{\phi}\tilde{L})\partial_{r'}\tilde{L}dr'$ \citep{Page_Thorne_1974a}. Here, $\dot{M}$ is the black hole's accretion rate, which is assumed to be constant, $\tilde{E}$ and $\tilde{L}$ are the disk's specific energy-at-infinity and specific angular momentum, respectively, and $r_I=3r_S$ is the radius of the innermost stable circular orbit (ISCO). The analytical integral expression for temperature can be approximated by 
	\begin{align}
		T(r)=T_0\left(\frac{r_S}{r}\right)^{\beta}\left(1-\sqrt{\frac{r_I}{r}}\right)^{\lambda},\label{eq:temp}
	\end{align}
	where $\beta$ and $\lambda$ characterize the power law behavior of the temperature. Here, the constant $T_0 [K]$ encloses the constants of the blackbody radiation temperature derived by \cite{Thorne_1974}. This behavior is based on analytical expressions for the surface temperature of matter in an accretion disk around a black hole with $\beta=3/4$ and $\lambda=1/4$ \cite{Shakura1973, Page_Thorne_1974a, Bhattacharyya_Thampan_Misra_Datta_2000}.
	
	The plasma density in the accretion disk can be obtained from the continuity equation $\partial_{r}(\sqrt{|g|}n\Gamma v^r)/{\sqrt{|g|}}=0$. To compute the value of the density, we approximate the orbits as spirals given by $r(\phi)=r_0e^{-\zeta\phi}$, where $\zeta$ is a function of $r$ but can be considered as a constant for very tightly bound orbits \citep{bhattacharjee_das_mahajan_20152}. We then take the limit $\zeta\to 0$, which vanishes $v^r$. The resulting plasma density satisfies
	\begin{align}
		n(r)=\frac{n_0}{\Gamma(r)} \left(\frac{r_S}{r}\right)^{3/2},
		\label{n}
	\end{align}
	where $n_0$ is a dimensional constant with units of [cm$^{-3}$].
	By replacing \eqref{eq:heat_a}, \eqref{eq:heat_b}, \eqref{eq:temp} and \eqref{n} in Eq.~\eqref{eq:lambdaheatflux}, we obtain that the only non-vanishing term is in the azimuthal direction
	\begin{eqnarray}
		\Lambda^{\phi}(r)=-\frac{\kappa_0 T_0}{qc^2}\frac{v^{\phi}(r)}{n(r)}\frac{(T(r)/T_0)^{7/2}}{r^2}h(r),\label{eq:Lambda_final}
	\end{eqnarray}
	where
	\begin{align}
		h(r)=&-\left(2-\frac{\Gamma^2(r)}{2}-\frac{7}{2}\left[\beta-\frac{\lambda}{2}d(r)\right]\right)k(r)\nonumber\\
		&+\frac{\lambda }{4}d(r)\left(d(r)+1\right)+\left(\frac{1}{2}+\beta-\frac{\lambda}{4}\big[3+d(r)\big]\right) \frac{r_S}{r} ,\label{eq:h}\\
		k(r)&=-\beta+\frac{\lambda}{2}d(r)+\left(\frac{1}{2}+\beta-\frac{\lambda}{2}d(r)\right) \frac{r_S}{r} ,\label{eq:f}\\
		d(r)=&\frac{\sqrt{{r_I}/{r}}}{1-\sqrt{{r_I}/{r}}}.\label{eq:g}
	\end{align}

	Let us note that (\ref{eq:Lambda_final}) is unavoidably singular at the black hole singularity and at $r=r_{ph}$, due to the appearance of the Lorentz factor both in (\ref{eq:h}) and in the ratio $v^{\phi}(r)/n(r)$. However, the function $d(r)$ that arises as a consequence of the the factor $(1-{{r_I}/{r}})^{\lambda}$ in (\ref{eq:temp}), implies that the vorticity obtained from (\ref{eq:vorticity}) will be singular at the ISCO for any $\lambda<{6}/{7}$, for which the domain of the final solution will be limited to that of stable orbits only.
	
	The curl of vector \eqref{eq:Lambda_final} does not vanish in general, and then the heat flux drive \eqref{heatfluxdrive} does indeed produce an initial temporal variation of the generalized vorticity in the polar direction.
	
	Since we are interested in studying the generation of magnetic fields, it is useful to explicitly separate the generalized vorticity into the magnetic field, which will be initially null, and the fluid vorticity, which will be non-zero in our model. Because of the initial lack of electromagnetic fields in \eqref{eq:Lorentz}, we consider an initial equilibrium state in which the time derivatives of the fluid velocity, particle density, temperature, and pressure are zero. The second term in the left-hand side of \eqref{eq:vorticity} is zero for our initial conditions, and we are left with the following equation for the time evolution of the magnetic field
	\begin{align}
		r\partial_t B^{\theta}&=-c\alpha(r)[\partial_r\Lambda^{\phi}+\frac{1}{r}\Lambda^{\phi}]\nonumber\\
		&\quad+r\Big(\frac{m\alpha(r)}{qc}\Big)[\partial_r(\Gamma v^{\phi}\partial_t f)+\frac{1}{r}(\Gamma v^{\phi}\partial_t f)].
	\end{align}
	Here, the magnetic field is multiplied by $r$ to obtain a result in magnetic field units since the choice of a spherical coordinate basis implies the angular components of vectors are divided by $r$.
	
	Because of the inclusion of a heat flux, the system will be unstable as energy will be injected into it. The time derivative of the enthalpy related quantity $f$ can be obtained from the energy equation $U_{\mu}\nabla_{\nu}T^{\mu\nu}=0$, which yields $\partial_t f=[-({1}/{\Gamma})\nabla_{\nu}q^{\nu}+({\Gamma v^2}/{rc^2})q^r]/mn(r)$.
	
	Following \cite{asenjo_mahajan_qadir_2013}, if the magnetic field is zero at the initial instant $t=t_0$, its time evolution can be approximated to first order as $\partial_t B^{i}(t)\big|_{t=t_0}^{t=\varsigma}\approx B^{i}(\varsigma)/\varsigma$, where $\varsigma={r}/{|\vec{v}|\alpha}\approx{1}/{v^{\phi}\alpha}$ is the timescale for a linear generation of a magnetic seed. Considering initially stable orbits and, therefore, a total axial symmetry of the system, then the baroclinical Biermann battery and the general relativistic drive both vanish \citep{asenjo_mahajan_qadir_2013}. Under these conditions, a careful treatment of (\ref{eq:vorticity}) yields the magnetic seed that is generated in a time interval $\varsigma$
	
	\begin{align}
		r B^{\theta}(t_0)\approx&-\frac{c}{v^{\phi}}[\partial_r\Lambda^{\phi}+\frac{1}{r}\Lambda^{\phi}]\nonumber\\
		&+r\Big(\frac{m}{qcv^{\phi}}\Big)[\partial_r(\Gamma v^{\phi}\partial_t f)+\frac{1}{r}(\Gamma v^{\phi}\partial_t f)]. \label{eq: vorticity_final}
	\end{align}
	
	\begin{figure}[h!]
		\centering
		\includegraphics[width=0.5\textwidth]{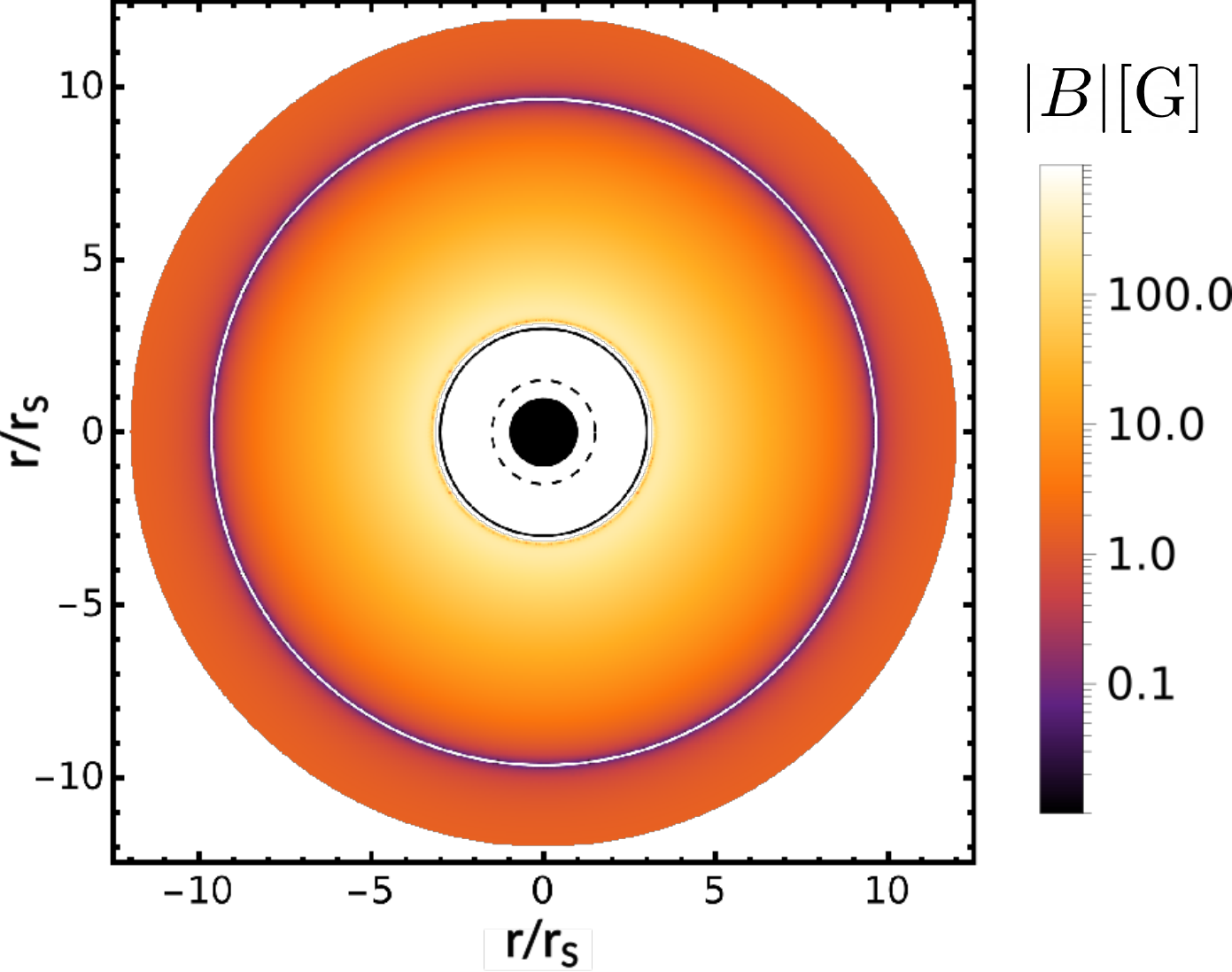}
		\caption{Magnitude of the magnetic field produced in the equatorial plane in a time interval $\varsigma$. The shaded area represents the inside of the horizon, the dashed line indicates the photosphere, the solid black line is the ISCO, and the solid white line is the radius at which the generated magnetic field passes through zero and inverts its sign.
		}
		\label{fig:map_vorticity}
	\end{figure}

	Equation \eqref{eq: vorticity_final} is solved numerically for $\beta=3/4$, $\lambda=1/4$, which indicates a black-body radiation-dominated temperature, as well as $n_0=10^{10} \textnormal{cm}^{-3}$, $T_0=10^{6} \textnormal{K}$ \citep{Bhattacharyya_Thampan_Misra_Datta_2000,Narzilloev_2022} and $\kappa_0=10^{9} \textnormal{g}\cdot\textnormal{s}^{-3}\cdot\textnormal{K}^{-1}$, following  Ref.~\cite{Meyer-Hofmeister_Meyer_2006}. The charge and mass are taken to be those of the electron. We focus on solutions for $r<20r_s$, as this is the region where gravity is strong and the effects of the generalized vorticity batteries are most important. For this choice of the parameters $\beta$ and $\lambda$, the poloidal magnetic field generated in a time interval $\zeta$ diverges at $r_I$ and becomes 0 at $r_{0B}/r_S=9.6$. Similarly, the time derivative of the enthalpy-weighted fluid vorticity $\partial\omega/\partial t$, where $\vec{\omega}=\vec{\nabla}\times(f\Gamma\vec{v})$, vanishes at $r_{0\omega}/r_S=9.8$. The heat flux battery $\Xi_{q}$ becomes zero at $r_{0q}/r_S=11.9$.
	
	The vanishing radii for $B$, $\partial \omega/\partial t$, and $\Xi_q$ are strongly dependent on $\beta$ and $\lambda$, which in this work are chosen according to values commonly associated with black hole accretion disks. In general, an increment in $\lambda$ will increase the values of $r_{0B}$, $r_{0\omega}$ and $r_{0q}$, while an increase in $\beta$ has the opposite effect.  
	
	Fig.~\ref{fig:map_vorticity} displays the strength in Gauss of the magnetic field generated in the time interval $\zeta$ in the equatorial plane as a function of $r/r_S$. 
	
	\begin{figure}
		\centering
		\includegraphics[width=0.5\textwidth]{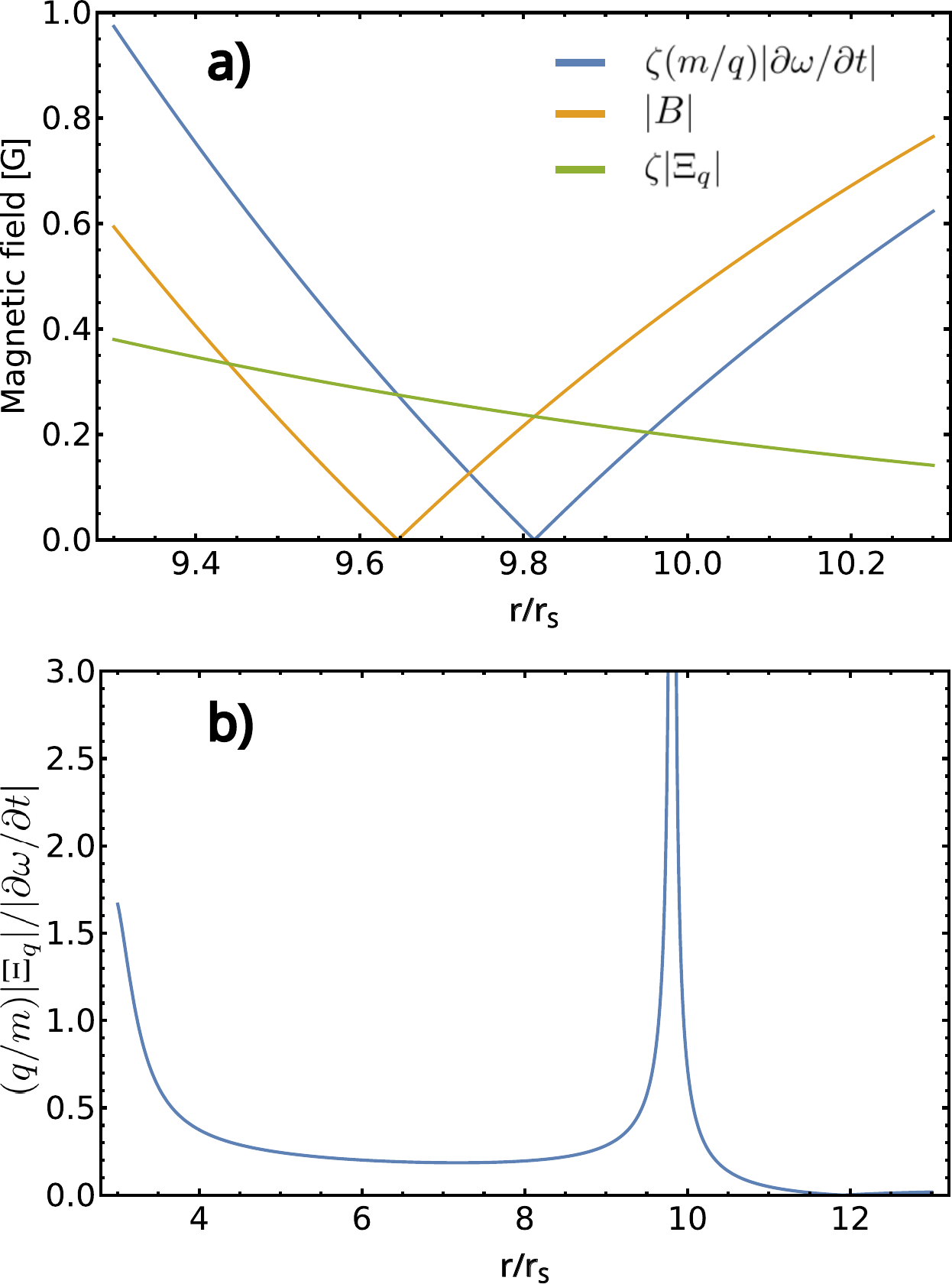}
		\caption{ (a) Magnitude of the linear-time evolution of the magnetic field $|B|$, heat flux battery $\zeta|\Xi_q|$  and fluid vorticity $\zeta(m/q)|\partial\omega/\partial t|$ as functions of $r/r_S$. (b) Ratio of the strengths of the heat flux battery and time evolution of the fluid vorticity as a function of $r/r_S$.
		} 
		\vspace{0ex}
		\label{fig:vorticity2}
	\end{figure}
	
	Fig.~\ref{fig:vorticity2}(a) displays the amplitude of the magnetic field $|B|$ generated in a time interval $\zeta$, as well as the absolute values of the linear-time evolution of the heat flux battery $\zeta|\Xi _q |$ and fluid vorticity $\zeta(m/q)|\partial \omega/\partial t|$ in magnetic field units (first and second term in Eq.~\eqref{eq: vorticity_final}, respectively), all as functions of $r/r_s$. Fig.~\ref{fig:vorticity2}(b) depicts the ratio $|\Xi_q|/(m/q)|\partial\omega/\partial t|$ between the heat flux battery and the time evolution of the fluid vorticity. Both figures are displayed to show that, since $r_{0\omega}\neq r_{0q}$, there is a domain for which the magnetic field generated in an initial time interval $\zeta$ will be driven by the heat flux battery. This domain is characterized by having $|\Xi_q|/(m/q)|\partial\omega/\partial t|>1$, and for this case study, the latter inequality is satisfied for $3<r/r_S<3.2$ and $9.6<r/r_S<9.9$. As with the vanishing radii, the location of this domain is highly dependent on the particular choice of the parameters $\beta$ and $\lambda$ in \eqref{eq:temp}.
	
	\begin{figure}[H]
		\centering
		\includegraphics[width=0.45\textwidth]{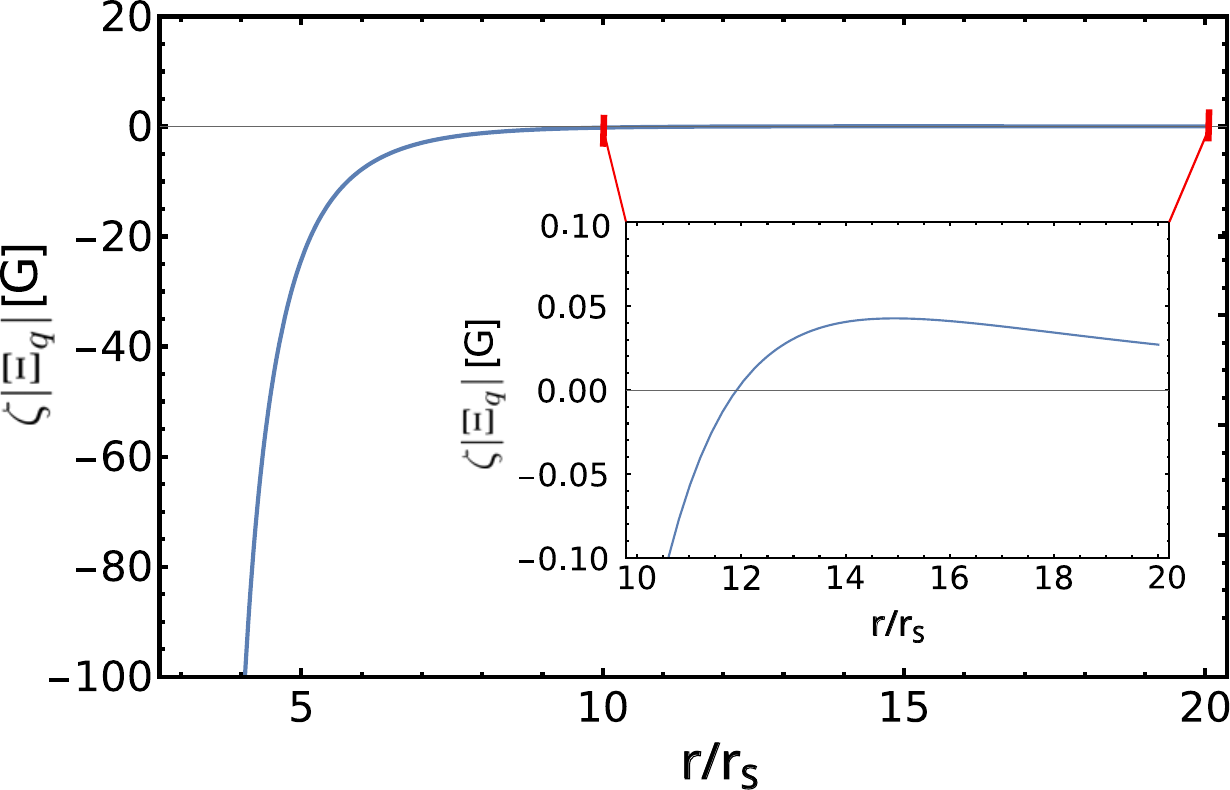}
		\caption{Generalized vorticity generated by the heat flux battery in a time interval $\zeta$ as a function of $r/r_S$.
		}
		\label{fig:vorticity3}
	\end{figure}
	
	Fig.~\ref{fig:vorticity3} displays the behavior of the heat flux battery $\Xi_{q}$ as a function of $r/r_S$. As previously discussed, this quantity diverges at $r_I$. The value $r_{0q}$ at which $\Xi_q$ vanishes is where this battery switches its sign. For $r>r_{0q}$, $\Xi_q$ reaches a local maximum and vanishes as $r/r_S\to\infty$. The asymptotic behavior of this quantity is a consequence of the choice of the temperature profile in \eqref{eq:temp}.  
	\section{Discussion}
	
	In this article, we have shown that the inclusion of non-ideal thermodynamic effects in the description of a charged fluid orbiting a non-magnetized black hole contributes to the generation of a magnetic field in the system. In particular, if the system is axisymmetric and well described by Keplerian dynamics, only the novel heat flux battery $\Xi_q$ induces the time evolution of initially non-existent generalized vorticity since both the Biermann ($\Xi_B$) and the Relativistic ($\Xi_R$) batteries vanish because of this symmetry (see \cite{Khanna_1998b} and \cite{asenjo_mahajan_qadir_2013} for further discussion on the role of the symmetric properties of the plasma and geometry on the generation of magnetic fields). To our knowledge, this is the first time that non-ideal thermodynamic effects are considered in deriving a general-relativistic generalized vorticity equation for magnetic fields in plasmas. However, heat flux has been widely considered in general-relativistic hydrodynamics and black-hole accretion disk theory, both from the point of view of physics and astrophysics \cite{Page_Thorne_1974a,misner_thorne_wheeler_kaiser_2017,Meyer-Hofmeister_Meyer_2006}.
	
	By proposing a simple model of a thin plasma disk with no accretion and a conduction-dominated heat flux, we demonstrate that the heat flux battery $\Xi_q$ is the only initial source for the time evolution of the magnetofluid's generalized vorticity. In particular, if the system's magnetic field is initially zero, including the heat flux is directly responsible for generating a magnetic seed at small temporal scales. We show that the heat flux battery is the primary source of the magnetic field in the region $9.6<r/r_S<9.9$, which is where our assumptions regarding the model are most valid. Our model can be extended by considering more sophisticated velocity profiles that include disk accretion and by addressing the full non-linear time evolution of the magnetofluid vorticity equation. The role of accretion on the time evolution of the magnetic field has been discussed extensively in the context of the \textit{Cosmic Battery} \citep{Contopoulos_2015}.

		It is important to note that the use of the \cite{Spitzer_Härm_1953} thermal conductivity is only valid in an electron-proton plasma. We are thus working over that supposition \textit{ab initio}. The heat flux discussed in this article is driven mainly by electrons andresults from the gradients of the red-shifted radiation temperature $\bar{T}=\alpha T$ of the disk in \eqref{eq:heat_b}. The coupling between the plasma velocity and the energy flux contributes to a non-zero $\Lambda^{\phi}$ in \eqref{eq:lambdaheatflux}, which gives rise to an azimuthal generalized electric field in \eqref{eq:gen-ef}. The generalized electric field, in turn, can be related to an azimuthal generalized current through an analogous Ohm's law in the unified magnetofluid formalism, which draws parallelisms to the \textit{Cosmic Battery} proposed by \cite{Contopoulos_1998}. In the \textit{Cosmic Battery}, the radiation temperature is also the source of an azimuthal current through the aberration of the radiation force acting on the electrons due to the azimuthal motion of the plasma (see \cite{Koutsantoniou_2014} for a detailed discussion).

		The final results shown in this article depend strongly on the particular choices of the velocity and temperature profiles and the heat flux considered in our model. The underlying physical consequences of these results, however, are far-reaching. Eqs.~(\ref{eq:lambdaheatflux}) and ~(\ref{eq:vorticity}) show that the only restriction for this new generalized vorticity battery to exist is that $\nabla_j \Lambda_k \neq 0$, which can be achieved by a wide variety of functions describing the velocity profile and heat flux within the accretion disk. Thus, our results are not limited to the particular magnetofluid configuration treated in this article as an example. On the contrary, this general result should be applied to any study that explores the generation of magnetic fields via a Biermann battery-like mechanism in both classical and relativistic regimes where energy fluxes are present, whether this flux is due to convection, radiation, conduction, or even a particle flux through the lens of the mass-energy equivalence \citep{landau1959fm}. This implies that our results can be further extended by including other possible sources of magnetic field generation, as well as black hole rotation, accretion, and explicit time dependence of thermodynamic and hydrodynamic quantities.

		\begin{acknowledgements}
			We thank Prof. Kinwah Wu (UCL-MSSL) for helpful discussions. N.V.S. is supported by ANID, Chile, through National Doctoral Scholarship N° 21220616. F.A.A thanks to FONDECYT grant No. 1230094 that partially supported this work. P.S.M is supported by FONDECYT Grant No. 1240281, and by the Research Vice-Rectory of the University of Chile (VID) through Grant ENL08/23.
		\end{acknowledgements}
		
		\bibliography{bibliography}{}
		\bibliographystyle{aasjournal}

	\end{document}